# Active tuning of highly anisotropic phonon polaritons in van der Waals crystal slabs by gated graphene

Gonzalo Álvarez-Pérez[1,2,*], Arturo González-Morán[3,*], Nathaniel Capote-Robayna[3], Kirill V. Voronin[4], Jiahua Duan[1,2], Valentyn S. Volkov[4], Pablo Alonso-González[1,2,†], Alexey Y. Nikitin[3,5,†]

[1]*Department of Physics, University of Oviedo, Oviedo 33006, Spain.*
[2]*Center of Research on Nanomaterials and Nanotechnology, CINN (CSIC-Universidad de Oviedo), El Entrego 33940, Spain.*
[3]*Donostia International Physics Center (DIPC), Donostia-San Sebastián 20018, Spain*
[4]*Center for Photonics and 2D Materials, Moscow Institute of Physics and Technology, Dolgoprudny 141700, Russia*
[5]*IKERBASQUE, Basque Foundation for Science, Bilbao 48013, Spain.*
*\*These authors contributed equally to this work*
*†Corresponding author. Email: pabloalonso@uniovi.es; alexey@dipc.org*

**Phonon polaritons (PhPs) —lattice vibrations coupled to electromagnetic fields— in highly anisotropic media display a plethora of intriguing optical phenomena (including ray-like propagation, anomalous refraction, and topological transitions, among others), which have potential for unprecedented manipulation of the flow of light at the nanoscale. However, the propagation properties of these PhPs are intrinsically linked to the anisotropic crystal structure of the host material. Although in-plane anisotropic PhPs can be steered (and even canalized) by twisting individual crystal slabs in a van der Waals (vdW) stack, active control of their propagation via external stimuli presents a significant challenge. Here, we report on a technology in which anisotropic PhPs supported by biaxial vdW slabs are actively tunable by simply gating an integrated graphene layer. Excitingly, we predict active tuning of optical topological transitions, which enable controlling the canalization of PhPs along different in-plane directions in twisted heterostructures. Apart from their fundamental interest, our findings hold promises for the development of optoelectronic devices (sensors, photodetectors, etc.) based on PhPs with dynamically controllable properties.**

Phonon polaritons in anisotropic vdW materials allow confining light at the nanoscale and its propagation only along certain specific directions[1-4], thus providing a natural playground to control, route and direct light at the nanoscale. This is because PhPs in such highly anisotropic media exhibit hyperbolic dispersion within the so-called reststrahlen bands (RBs) –spectral regions between transversal and longitudinal optical phonons. Within the RBs, at least one component of the real-part dielectric permittivity tensor, $\hat{\varepsilon}$, is negative, leading to a high reflectivity. If two components of $\hat{\varepsilon}$ have opposite signs, the iso-frequency curve (IFC) of PhPs —a slice of the dispersion surface in momentum or $\boldsymbol{k}$-space, being $\boldsymbol{k} = (k_x, k_y, k_z)$ the wavevector, by a plane of a constant frequency, ω— describes an open hyperbola. The landmark feature of hyperbolic PhPs (HPhPs) is their ray-like propagation inside the crystal, which can lead to out-of-plane hyper-focusing, as demonstrated for hexagonal boron nitride (h-BN)[5,6], or to dramatically enhanced photothermoelectric effects, as recently reported for heterostructures made of h-BN and graphene[7]. Particularly, HPhPs in some vdW crystals, such as alpha-phase molybdenum trioxide[3,4] (α-MoO$_3$) or alpha-phase vanadium pentaoxide[8] (α-V$_2$O$_5$), have their hyperbolic IFCs in the plane parallel to the faces of the vdW crystal slab (e.g., in the $k_x, k_y$ plane), thus showing in-plane ray-like propagation. This unique characteristic can result in exciting optical phenomena, such as in-plane canalization[9-13], or sub-diffractional planar hyper-lensing[14] and hyper-focusing[15,16]. However, dynamical control of the propagation properties of in-plane HPhPs, crucial for their potential implementation in optical and optoelectronic applications, such as tunable photodetection or sensing, has so far remained elusive.

Here, we demonstrate that in-plane HPhPs in vdW crystal slabs can be effectively manipulated by incorporating a gated graphene layer. Although our general theoretical results are valid for any biaxial slab (with all the main elements of its permittivity tensor being different $\varepsilon_x \neq \varepsilon_y \neq \varepsilon_z$), and hence for the less general case of uniaxial and isotropic slabs, we illustrate our idea for molybdenum trioxide (α-MoO₃), a biaxial vdW crystal which has recently received significant attention in optics and material science[3,4,9-12,15,16]. Due to coupling between HPhPs in α-MoO₃ slabs and graphene plasmon polaritons (GPPs)[17], hybrid polaritons in graphene/α-MoO₃ heterostructures become actively tunable via electrostatic gating of the graphene layer[17-20]. Importantly, we show the possibility of an active control of polariton canalization in twisted slabs of α-MoO₃. Furthermore, hybrid polaritons in graphene/α-MoO₃ heterostructures exhibit low ohmic losses, thus inheriting the long propagation length and lifetime from PhPs in α-MoO₃.

Figure 1a shows a schematic of the proposed device, in which a gated graphene layer is placed on top of an α-MoO₃ slab of thickness $d$. The dispersion of polaritons in such graphene/α-MoO₃ heterostructure can be analytically derived by generalizing the dispersion relation of polaritons in biaxial slabs surrounded by two semi-infinite media in the high-momenta approximation[21,22]:

$$k(\omega) = \frac{\rho}{d}\left[\arctan\left(\frac{\varepsilon_1 + 2i\alpha_g k/k_0}{\varepsilon_z}\rho\right) + \arctan\left(\frac{\varepsilon_3}{\varepsilon_z}\rho\right) + \pi l\right], \quad l = 0, 1, 2 \ldots, \quad (1)$$

where $k$ is the in-plane wavevector ($k^2 = k_x^2 + k_y^2$), $k_0 = \omega/c$ is the free-space wavevector, $\varepsilon_1$ and $\varepsilon_3$ are the dielectric permittivities of the superstrate and substrate, respectively, $d$ is the thickness of the slab, $\rho = i\sqrt{\varepsilon_z/(\varepsilon_x \cos^2\varphi + \varepsilon_y \sin^2\varphi)}$ with $\varphi$ being the angle between the x axis and the in-plane wavevector, and $l$ is the counting number. $\alpha_g$ is the 2D conductivity of graphene[23], which we will take at ambient conditions all throughout the work. Eq. (1) is an implicit expression, (since $k$ appears both in the left part and in the argument of the arctan), and can be solved numerically. The different polaritonic modes in the heterostructure, described by Eq. (1), present hybridization between HPhPs in α-MoO₃ slab and GPPs, with $l$ characterizing the quantization of the electromagnetic field of the mode in the $z$ direction, in such a way that the higher the $l$, the stronger confinement of the modes and the shorter wavelength and propagation length. An instructive way to represent the dispersion of polaritonic modes is to plot the imaginary part of the corresponding Fresnel reflection coefficient, $\text{Im}[r_p(\omega, k)]$, obtained from transfer-matrix calculations[24], as a function of frequency and wavevector (Fig. 1b). In our calculations the thickness of the α-MoO₃ slab is set to $d = 120$ nm (thus being in the range of typical thicknesses in recent near-field experiments), while the relaxation time of charge carriers in graphene is set to $\tau_{\text{rel}} = 0.5$ ps. Although this value of $\tau_{\text{rel}}$ is typical for encapsulated graphene used in optical experiments at ambient conditions[25,26], we do not consider the encapsulating h-BN layers in our calculations since they do not introduce any significant changes, and particularly, in-plane effects to the PhPs dispersion. For clarity, the results for $\text{Im}[r_p(\omega, k)]$ at different Fermi levels of the graphene layer, $E_F$ (or, equivalently, different gate voltages) are superimposed on the same colorplot in Fig. 1b. The bright maxima correspond to the dispersion of polaritonic modes in the heterostructure. At the same time, the dashed lines render the lowest-order modes ($l = 0$) calculated according to Eq. (1). On the one hand, the perfect matching between the maxima of the colorplot and the curves proves the validity of our simple analytical approximation. On the other hand, we clearly see that the dispersion relation of the mode of the same index $l$ is strongly dependent upon $E_F$. Namely, the dispersion along both the [100] and [001] crystal directions become steeper as $E_F$ increases from 0 to 0.7 eV, so that the momentum of polaritons decreases with $E_F$ at a fixed frequency. Hence, the tuning of $E_F$ provides an effective method to control the sub-diffractional confinement of PhPs in heterostructures made of graphene and a biaxial polaritonic slab, allowing to do so, in addition, in an active way (e.g., by an applied gate voltage).

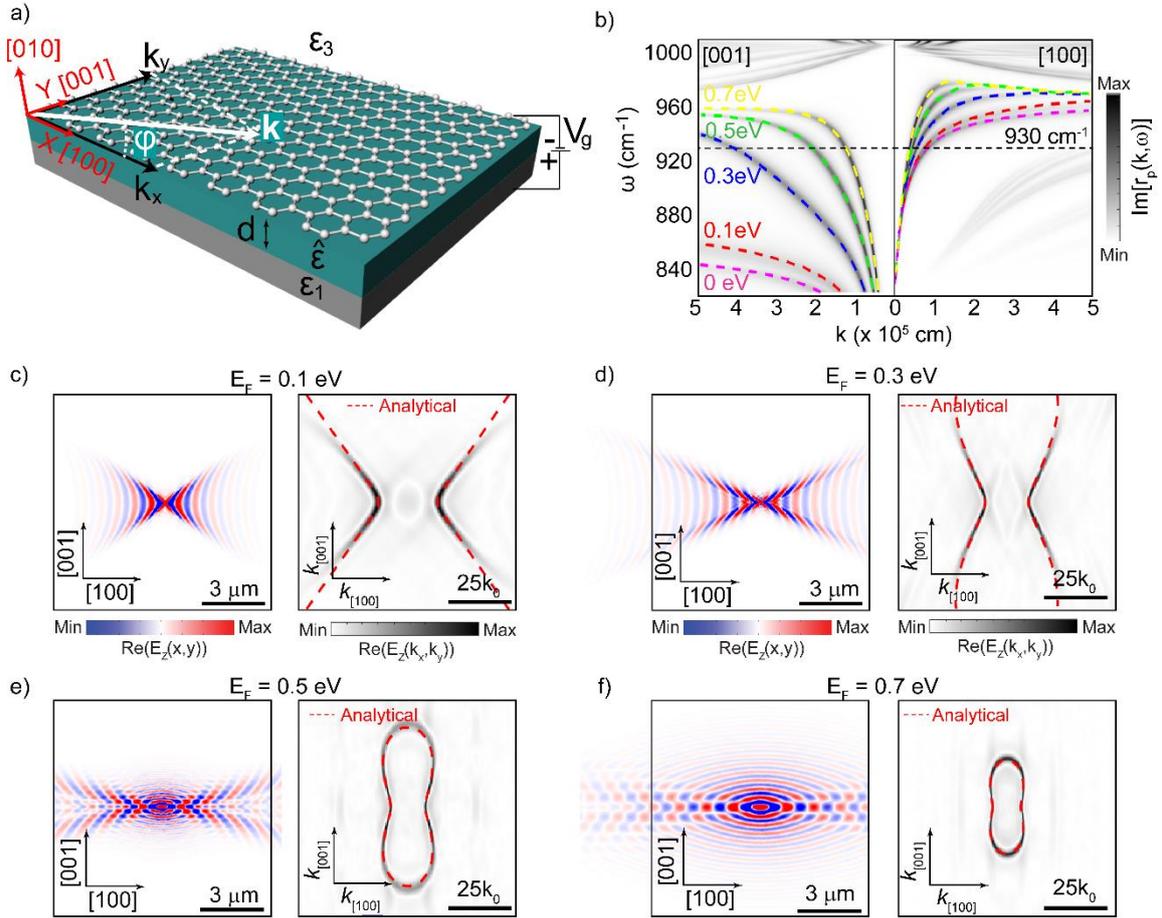

**Fig. 1 | Active tuning of hybrid polaritons in a graphene/α-MoO₃ heterostructure. (a)** Schematics of an α-MoO₃ biaxial slab with a gated graphene sheet placed on top. **(b)** Dispersion of hybrid polaritons in α-MoO₃/graphene for different representative values of $E_F$ = 0.1, 0.3, 0.5 and 0.7 eV. The false color plot displays transfer-matrix calculations showing the imaginary part of the Fresnel reflection coefficient, while the analytical dispersions are depicted as dashed lines. **(c-f)** Simulated electric field spatial distribution (left panels) and corresponding numerical and analytical polaritonic IFCs (colorplots and dashed red lines in the right panels, respectively) of polaritons for $E_F$ = 0.1, 0.3, 0.5 and 0.7 eV, showing a topological transition

from open hyperbolas (c, d) to closed curves (e, f) and canalization (f). The incident frequency is fixed to $\omega_0 = 930$ cm$^{-1}$.

To study in more detail the polaritonic features as a function of $E_F$ in graphene/α-MoO$_3$ heterostructures, we calculate the spatial distribution of the vertical component of the electric field, $E_z(x, y)$, created by a vertical electric point dipole at different $E_F$. For the sake of simplicity, we perform our calculations at a representative frequency $\omega_0 = 930$ cm$^{-1}$ (shown by a horizontal dashed line in Fig. 1b) belonging to the hyperbolic RB of α-MoO$_3$, which spans the spectral region between $\omega_0 = 821.4$ and $963.0$ cm$^{-1}$ (RB$_2$)[27]. For a Fermi level $E_F = 0$ eV, i.e., the case without graphene, the near-field distribution features concave wavefronts, as typically observed for HPhPs in α-MoO$_3$[3,4]. For $E_F = 0.1$ eV, PhPs also propagate within a hyperbolic sector centered along the [100] direction, which, however, presents now a lower angle (Fig. 1c, left panel). By increasing $E_F$ to 0.3 and 0.5 eV, such propagating sector becomes much narrower (left panels in Figs. 1d and 1e, respectively), indicating a further reduced hyperbolic angle, while, on the other hand, the PhPs wavelength increases. Interestingly, in the case of $E_F = 0.5$ eV, additional elliptical wavefronts appear in the neighbourhood of the origin. Excitingly, when reaching $E_F = 0.7$ eV, the propagating hyperbolic sector virtually converts into a line (Fig. 1f, left panel), so that the PhPs appear to propagate in their majority only along one specific direction (canalization regime). Again, elliptical wavefronts are also present in the electric field distribution. These results unambiguously demonstrate that the $E_z$ distribution of polaritons in a graphene/α-MoO$_3$ heterostructure exhibits $E_F$-dependent wavefronts, thus corroborating that such hybrid polaritons inherit the extremely directional nature of HPhPs in α-MoO$_3$ while gaining a unique virtue of GPPs: tunability via $E_F$.

In order to further characterize the propagation properties of hybrid PhPs in a graphene/α-MoO$_3$ heterostructure, we analyze the polaritonic IFCs in momentum space. To that end we perform the Fourier-transform (FT) of the near-field $E_z(x, y)$ simulations in Figs. 1 c-f (represented by colorplots Figs.1 c-f, right panels). The FTs for $E_F = 0$ eV and 0.1 eV (Fig. 1c, right panel) show open hyperbolas, as expected from the concave shape of the wavefronts in the field distributions, and in excellent agreement with our analytical calculations (dashed red curves). For $E_F = 0.3$ eV (Fig. 1d, right panel), the opening angle of the hyperbola (given by the asymptote $|\tan(k_x/k_y)| = \sqrt{-\varepsilon_y/\varepsilon_x}$) increases, consistently with the field distribution, still showing concave wavefronts (Fig. 1d, left panel). Conversely, for $E_F = 0.5$ eV (Fig. 1e, right panel), the FT presents a closed curve, revealing both a hyperbolic-like shape IFC along the [100] direction and an elliptical-like shape along the [001] direction. Finally, for $E_F = 0.7$ eV, the FT reveals a closed IFC that covers a smaller area (Fig. 1f, right panel), consistently with the smaller momenta of GPPs for larger $E_F$.. Note that the IFC is not a perfectly closed ellipse owing to the circular wavefronts around the origin, yet it exhibits a strong flattening along the [100] direction, explaining the canalization of PhPs along this direction. We note that all numerical IFCs (colorplots) in Figs. 1c-f show an excellent agreement with the analytical curves given by Eq. (1) (red dashed curves), thus reconfirming its versatility. Importantly, apart from the strong dependence of the IFCs upon $E_F$, their evolution unveils a topological transition. Such transition can be characterized by a topological quantity, namely the genus of the IFC, which describes the number of holes in the curve. The genus of the IFCs in Figs. 1c, d, (left panels) is equal to 0, while for the IFCs in Figs. 1e, f, (left panels) it is equal to 1. The transition frequency $\omega_T$ is highly dependent upon $E_F$, that is for a given $\omega_T$ we can find a certain $E_F$ at which canalization of polaritons takes place. In particular, at $\omega_0 = 930$

cm$^{-1}$, the transition takes place at $E_F = 0.32$ eV, and it is well captured by our analytical equation. From a practical perspective, these results show that the topology of the polaritonic IFCs in a graphene/α-MoO$_3$ heterostructure can be actively controlled by electrical means. Note that up to now such a dramatic modification of the topology of the IFCs for in-plane anisotropic PhPs has only been obtained via coupling with polar substrates[28,29] or by fabricating twisted stacks of the polaritonic material[9-12], i.e., by approaches precluding any direct active control.

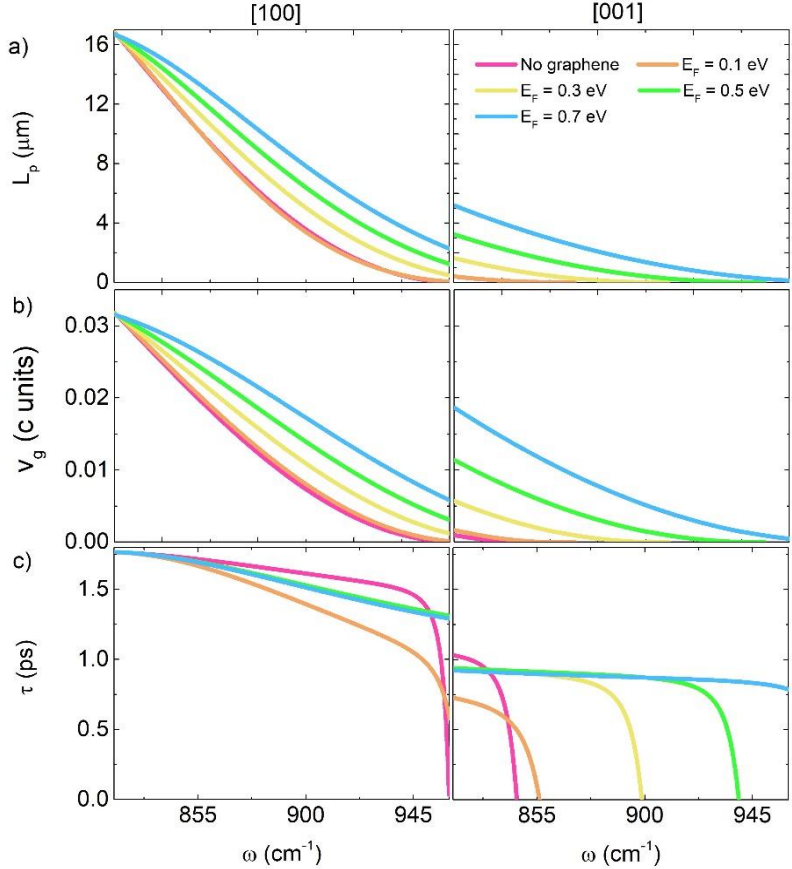

**Fig. 2 | Propagation properties of hybrid polaritons in a graphene/α-MoO$_3$ heterostructure.** (a) Analytical propagation length $L_p$, (b) group velocity $v_g$ and (c) lifetime $\tau$ as a function of frequency for different Fermi energies $E_F$ = 0.1, 0.3, 0.5, and 0.7 eV, together with the calculation for the case without graphene. The left (respectively, right) panel shows the calculations for polaritons propagating along the [100] (respectively, [001]) direction.

To gain further insights into the properties of hybrid polaritons in a graphene/α-MoO$_3$ heterostructure, we quantitatively characterize their crucial parameters: propagation length, group velocity and lifetime as a function of the incident frequency $\omega$ and Fermi energy $E_F$. Fig. 2a shows the propagation length, $L_p = 1/\text{Im}(k)$, as a function of frequency, $\omega$. $L_p$ monotonically decreases with $\omega$ for polaritons propagating along both the [100] and [001] crystal directions (left and right panels of Fig 2a, respectively), which can be explained by the higher confinement of polaritons with increasing $\omega$. In contrast, $L_p$ increases with $E_F$, with this tendency being more dramatic for the [001] crystal direction, along which $L_p$ increases by a factor of 10 as $E_F$ increases from 0.1 to 0.7 eV (as can be seen e.g. at $\omega_0$ = 820 cm$^{-1}$). Similarly, the group velocity, $v_g = \partial\omega/\partial k$, increases with $E_F$ (Fig. 2b) while decreasing with $\omega$, showing the same trend along both in-plane crystal directions. Having calculated both $L_p$ and $v_g$, the polariton lifetime, $\tau$, can be easily analysed following the relation $\tau = L_p/v_g$. In contrast to $L_p$ and $v_g$, the dependence of $\tau$ upon $\omega$ is rather flat (Fig. 2c), reaching appreciable values of ~2 ps along the [100] crystal direction and of ~ 1 ps along the [001] direction. Importantly, although $\tau$ shows a steep drop along the [001] direction, which blueshifts with increasing $E_F$, the maximum value of $\tau$ is virtually independent upon $E_F$ (remaining high up to the dropping frequency even for large values of $E_F$) in the frequency range analysed. In fact, the spectral range where $\tau$ takes an appreciable value along the [001] direction (~ 1 ps, i.e., as long as when considering the slab without graphene) increases with $E_F$. Hence, the polaritonic propagation properties in heterostructures made of graphene and α-MoO$_3$, such as propagation length and lifetime, are not undermined by the active tuning capability, but instead can be substantially improved. As a result, the tunable propagating hybrid polaritons in graphene/α-MoO$_3$ heterostructures preserve the low-loss character of PhPs in α-MoO$_3$, which is crucial for applications requiring long-lived light-matter interactions, such as e.g. mid-infrared sensing or strong coupling between polaritons and molecules.

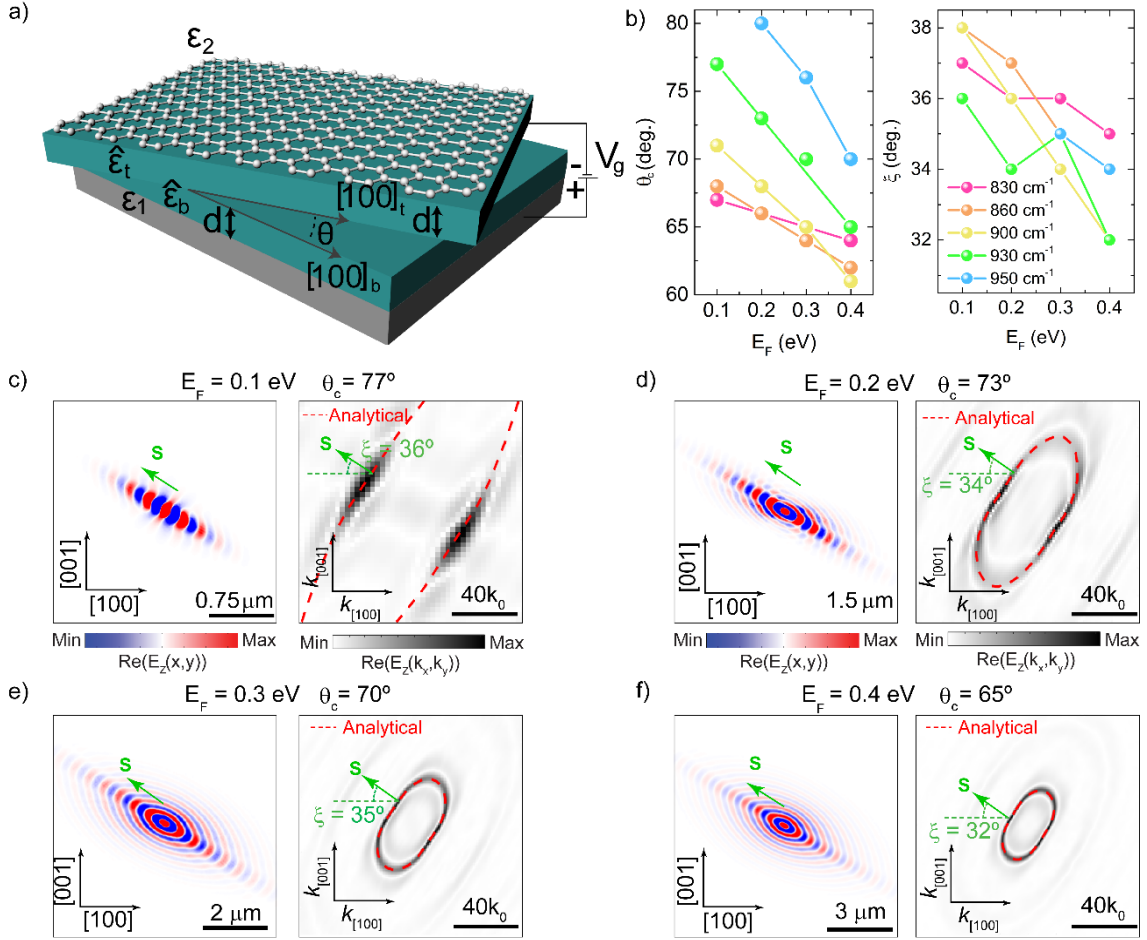

**Fig. 3 | Active tuning of hybrid polaritons in twisted vdW stacks.** (a) Schematics of two α-MoO$_3$ slabs superimposed with their in-plane crystalline directions rotated by a critical twist angle θ$_c$ and with a gated graphene layer placed on top. (b) Dependence of θ$_c$, and the in-plane angle at which canalization takes place, ξ, as a function of $E_F$ for different ω. (c-f) Simulated electric field spatial distribution (left panels) and corresponding numerical and analytical polaritonic IFCs (color plots and dashed blue lines in the right panels, respectively) of PhPs propagating in twisted α-MoO$_3$ stacks with a gated graphene layer placed on

top for $E_F$ = 0.1, 0.2, 0.3 and 0.4 eV, and $\theta_c$ = 77º, 73º, 70º and 65º. The incident frequency is fixed to ω = 930 cm$^{-1}$.

Going one step further, we also study active manipulation of polaritons by considering vdW stacks formed by two mutually twisted α-MoO$_3$ slabs covered by a graphene layer. Recent studies have shown that PhPs in two twisted α-MoO$_3$ slabs display canalization due to the emergence of a topological transition in the IFC[9-12]. However, in such structures the twisting angle at which the canalization of PhPs takes place (as well as the in-plane canalization angle itself) is predefined by the thicknesses of the slabs and the incident frequency. Namely, for a given thickness and incident frequency, canalization of PhPs in twisted structures can only be achieved for one specific twisting angle and one specific in-plane direction of propagation, thus severely limiting the tunability of these directional polaritons. Here, we introduce an additional degree of freedom for active tuning of canalized polaritons in twisted vdW stacks by adding a gated graphene layer. In particular, we consider a vdW heterostructure formed by two 50-nm-thick slabs of α-MoO$_3$ (the choice of this representative thickness does not affect the generality of our results) twisted by an angle θ with respect to each other and a graphene layer placed on top (schematics in Fig. 3a). In order to illustrate the propagation of polaritons in such stack, we calculate the spatial distributions of the electric field generated by a point dipole source for different $E_F$ (Figs. 3c-f) at ω = 930 cm$^{-1}$, as well as the corresponding FTs (analogously to the case of a single α-MoO$_3$ slab shown in Fig. 1).

For the case $E_F$ = 0 (i.e., no graphene) and by monitoring the spatial electric field distributions at different twist angles θ, we find that only one critical twist angle, $\theta_c$ = 75º, yields a canalization regime in which the excited polaritons propagate in a rather narrow angular sector. The corresponding direction along which the canalized polaritons propagate forms an angle $\xi$ = 41º with the $x$ axis. Interestingly, by adding a slightly doped graphene layer ($E_F$ = 0.1 eV), such critical angle decreases its value to $\theta_c$ = 77º, as observed in Fig. 3c (left panel). The corresponding FT (colorplot in Fig. 3c, right panel) shows a highly flat IFC, matching well the corresponding analytical result (red dashed lines). Due to the flatness of the IFC, the Poynting vectors (oriented perpendicularly to the IFC and denoted as $S$) of the plane waves with different wavevectors are parallel, forming an angle $\xi$ = 33º with the $x$ axis (Fig. 3c, right panel), which reconfirms that the polaritons propagate in the canalization regime. By increasing the doping of the graphene layer ($E_F$ = 0.2 eV), the critical twist angle decreases to $\theta_c$ = 73º and the canalized polaritons propagate within a narrower angular sector (Fig. 3d, left panel), corresponding to $\xi$ = 35º (Fig. 3d, right panel). Interestingly, some elliptical wavefronts appear close to the origin. The emergence of these wavefronts becomes clear after analysing the corresponding IFC (colorplot and red dashed lines in Fig. 3c, right panel), which although still giving rise to parallel Poynting vectors along its flat part, as expected for a canalized regime, also shows an elongated ellipse-like closed curve. As we further increase $E_F$ up to 0.3 eV (Fig. 3e), the critical twist angle at which canalization takes place further decreases to $\theta_c$ = 70º. For these parameters, the near field pattern reveals canalized polaritons with longer wavelengths and featuring elliptical wavefronts in the neighbourhood of the origin (Fig. 3e, left panel). Accordingly, the corresponding FT reveals a smaller closed IFC, which is flattened at an angle $\xi$ = 35º (Fig. 3e, right panel). By increasing $E_F$ up to 0.4 eV (Fig. 1f, left panel), the canalization regime requires decreasing the critical twist angle to $\theta_c$ = 65º. This is in accordance with the numerical and analytical IFCs, which are flattened at an angle $\xi$ = 37º (Fig. 3f, right panel). These results clearly indicate that for any given Fermi energy $E_F$, it is possible to find a critical twist angle $\theta_c$ at which canalization emerges. Conversely, for a predetermined twist angle, $\theta$, we can actively achieve canalization of polaritons by properly matching $E_F$, simply varying it between 0.1 and 0.5 eV for a rather wide range of twist angles. The results shown in Fig. 3c-f for an incident frequency $\omega = 930$ cm$^{-1}$ are collected in Fig. 3b (left panel), along with analogous calculations for other incident frequencies $\omega$ within the reststrahlen band RB$_2$ of α-MoO$_3$[27], spanning the range between $\omega = 821.4$ and $963.0$ cm$^{-1}$. These results clearly reveal that canalization of polaritons in our actively-tunable stack can be attained for a rather wide range of twist angles, between 60º and 80º, by simply varying $E_F$ between experimentally attainable values of 0 and 0.5 eV. In addition, canalization can occur at different in-plane angles $\xi$. In fact, the results shown in the right panel of Fig. 3b reveal that the in-plane direction along which the hybrid polaritons are canalized can be actively manipulated in an angular sector of more than 12º by simply varying $\omega$ and $E_F$. Note that while all these calculations are performed for fixed slab thicknesses of 50 nm, the twist and canalization angles $\theta_c$ and $\xi$ depend dramatically on such thicknesses. Therefore, these angles could be further manipulated for other combinations of parameters. Altogether, these findings open the door for the manipulation of highly directional (canalized) PhPs in a robust and dynamic way for a wide range of frequencies, angles and Fermi energies.

In conclusion, we report a pathway through which hyperbolic PhPs supported by biaxial vdW slabs can be actively tunable by integrating a gated graphene layer. We have unveiled tunable topological transitions of the polaritonic IFCs (leading to canalization of polaritons) which are highly dependent on the Fermi level of charge carriers in graphene. Excitingly, the propagating properties of the resulting hybrid polaritons, such as propagation length or lifetime (key for applications requiring long-lived light-matter interactions, as e.g. strong coupling or mid-infrared sensing), are not undermined by the ohmic losses in graphene. In addition, our findings demonstrate that low-loss canalization of polaritons can be also actively tuned in twisted vdW heterostructures where the in-plane canalization direction can be controlled in a certain range of angles, opening the door to steer infrared light along specific directions on demand. Apart from their fundamental relevance, our findings provide a basis to design near-field optical experiments and hold promises for optoelectronic devices (sensors, photodetectors, etc.) based on PhPs with dynamically controllable properties.

**Author Contributions**

J.D., P.A.-G. and A.Y.N. conceived the study. G.Á.-P., A.G.-M., N.C.-R. and K.V.V. performed the analytical calculations and the numerical simulations. G.Á.-P., A.G.-M. and N.C.-R. were responsible for data analysis and curation. All the authors discussed and developed the interpretation of the results. A.Y.N. and P.A.-G. coordinated and supervised the work. G.Á.-P. and A.Y.N. wrote the original manuscript with input from the rest of authors.

**Notes**

The authors declare no competing financial interest.

**Acknowledgments**

G.Á.-P. acknowledges support through the Severo Ochoa Program from the government of the Principality of Asturias (grant no. PA-20-PF-BP19-053). K.V.V. and V.S.V. acknowledge financial support from the Ministry of Science and Higher Education of the Russian Federation (Agreement No. 075-15-2021-606). P.A.-G. acknowledges support from the European Research Council under starting grant no. 715496, 2DNANOPTICA and the Spanish Ministry of Science and Innovation (State Plan for Scientific and Technical Research and Innovation grant number PID2019-111156GB-I00). A.Y.N. acknowledges the Spanish Ministry of Science and Innovation (grants MAT201788358-C3-3-R and PID2020-115221GB-C42) and the Basque Department of Education (grant PIBA-2020-1-0014).